\documentclass[aps,twocolumn,floats,showpacs,prb]{revtex4}
\usepackage{amsmath,amssymb,graphicx,epsf,epstopdf,epsfig}

\newcommand{\etal}{{\it et al.}}

\begin{document}

\title{Spin Zeros and the Origin of Fermi Surface Reconstruction in the Cuprates} 

\author{M. R. Norman}
\affiliation{Materials Science Division, Argonne National Laboratory, Argonne IL 60439}

\author{Jie Lin}
\affiliation{Materials Science Division, Argonne National Laboratory, Argonne IL 60439}

\begin{abstract}
Two recent quantum oscillation studies find contradictory results concerning the existence 
of spin zeros - zeros of the oscillatory signal induced 
by Zeeman splitting of the Landau levels.
We discuss these experiments in light of calculations of the oscillations assuming a
spin density wave state.  We find that the lack of spin zeros reported in one of
the experiments is consistent with either hole or electron pockets in such a state, 
if the staggered moment is perpendicular to the external field.
An analysis for field directions near the planes might be able to differentiate between the two. 
On the other hand, if spin zeros exist as reported in the other experiment,
then the staggered moment would have to have a substantial longitudinal component.
We suggest several experiments to test whether this is indeed the case.
\end{abstract}
\date{\today}

\maketitle

Since the pioneering work of Doiron-Leyraud \etal,\cite{Doiron} quantum oscillation studies have
provided  new insights into the nature of the electronic ground state of the cuprates.  For overdoped
compounds, a large Fermi surface is observed,\cite{Tl2201} consistent with previous photoemission
studies \cite{RMP} 
and paramagnetic band structure calculations.\cite{Ole}
For underdoped compounds, this Fermi surface breaks up into smaller 
pockets.\cite{Doiron,SS,Audouard,SSPRB}  It is thought that this break up is due to a density
wave reconstruction of the Fermi surface, perhaps due to the formation of magnetic
stripes.\cite{LeBoeuf,Millis}
These findings have had a significant impact on our understanding of the cuprate phase
diagram,\cite{Subir} though debates continue about the relation of these small pockets to
the Fermi `arcs'  observed in zero field photoemission studies.

A key insight into the nature of this underdoped high field ground state  was recently provided 
by Sebastian \etal.\cite{SSPRL}  They looked at the oscillations as a function of the angle of the magnetic
field relative to the 
crystallographic $c$-axis.  If Zeeman splitting were present, then such a rotation study should find
certain angles, known as spin zeros, where the oscillation amplitude would go through zero, with
higher angles exhibiting a $\pi$ phase shift relative to lower ones.  Up to an angle near 60$^\circ$,
they did not observe this effect.  This was confirmed in a more extensive study.\cite{SSPRB}  As they suggest,
such a finding would be consistent with the presence of a spin density wave
state,\cite{LeBoeuf,Millis,Harrison} as we will elaborate on below.
Subsequent to this, though, another quantum oscillation study was done where zeros from two different
frequencies were found from fits to the data, with one near 40$^\circ$ and the other near 
50$^\circ$.\cite{Ramshaw}  Based on this, the authors suggest a paramagnetic ground state instead.

These results raise a general question about the nature of the high field ground state, and how much
the Zeeman splitting is actually reduced if this ground state is a spin density wave state.
Recently, Ramazashvili,\cite{Revaz} following an earlier study by Kabanov and 
Alexandrov,\cite{Kabanov} has shown on symmetry grounds \cite{Revaz2}
that for commensurate spin density wave ordering,
$Q=(\pi,\pi)$, the Zeeman splitting should be quenched for both electron and hole pockets, as they
are centered at the magnetic zone boundary.  This is under the assumption that the magnetic field
strength is large enough for the spins to be reoriented perpendicular to the field.
On the other hand, for the incommensurate (stripe)
case, only the hole pockets have this symmetry.  His suggestion was this finding could be used
to resolve the controversy of whether the observed oscillation signal originated from electron or
hole pockets.

In this Rapid Communication, we perform 
calculations for a magnetic stripe in the presence of
an external field.  We verify that in the configuration where the spin moments are transverse to the 
field, the hole pockets do not exhibit any Zeeman splitting.  On the other hand, we find that the
Zeeman splitting of the electron pockets is strongly reduced compared to that of a
paramagnetic ground state.  Implications of our results in regards to the quantum oscillation
experiments, and proposals for new experiments, will be discussed.

Our model assumes a linear spin density wave state.\cite{Millis}
Without an external field, the secular matrix
has dimension $N$, where the magnetic ordering wavevector is $Q=(1-2/N,1)\pi$.  Diagonal
elements of this matrix are of the form $\epsilon_{k+nQ}$ with $n$ running from 1 to $N$,
where $\epsilon_k$ is the paramagnetic dispersion.  Off-diagonal elements due to the stripe
order are of the form $V_s \delta_{n,n \pm 1}$ and $V_c \delta_{n,n \pm 2}$ where $V_s$
is the magnetic potential, and $V_c$ its charge harmonic (for simplicity, we ignore higher
harmonics).

In the presence of an external field, spin must be explicitly taken into account, leading to a secular
matrix of dimension $2N$. Taking the spin quantization axis to be along the field direction, 
the Zeeman splitting
entering into the diagonal terms would be $\frac{g}{2} \mu_B H \sigma_z$ where $g$ is the g factor,
$\mu_B$ the Bohr magneton, $H$ the external field, and $\sigma_z$ a Pauli matrix.
The term $V_c$ would have a $\sigma_0$ matrix associated with it instead.  
Ignoring small effects due to spin-orbit coupling except for their renormalization of $g$,
then only the relative orientation of the staggered moment and the field is important. Thus, $V_s$ 
becomes $V_s(\sigma_x\sin\phi+\sigma_z\cos\phi)$ where $\phi$ is the angle between the staggered 
moment and the field. For $\phi=0$, the staggered moment is longitudinal, and for $\phi=\pi/2$ it
is transverse.
We note that for both the longitudinal and transverse cases,
the secular matrix decomposes into two diagonal blocks with dimension $N$.
The `down' spin block can be obtained from the `up' spin block by
inverting the sign of $H$ (and $V_s$ in the longitudinal configuration).  For the transverse
case, the `down' spin eigenvalue spectrum is equivalent to the `up' spin one if $k$ is translated by $Q$.
For the longitudinal case, with $V_c$=0, the secular matrices are equivalent to the paramagnetic
case except for a shift of $\epsilon_k$ by $\pm \frac{g}{2}\mu_B H$ (and thus one has full Zeeman splitting).
The transverse case is more interesting; detailed calculations presented below find that the Zeeman 
splitting is strongly suppressed.
For each $H$, we determine the chemical potential to obtain the correct occupation number
(which we take to be $1/N$ holes).

In Fig.~1a, we show the Fermi surface for $N$=8 in the transverse configuration,
with an artificially large value of $H$ so as to emphasize the Zeeman splitting.
The effect of the Zeeman field is to split each
of the two hole pockets (which have equal area)
by shifting their centers relative to one another.  But the split pockets have the same area,
as they are related to one another by the symmetries of the underlying Hamiltonian. 
This is associated with the fact that the Zeeman splitting is identically zero at the 
paramagnetic orbit centers,
as previously derived by Ramazashvili,\cite{Revaz} which he denotes as points $S \equiv
([2n+1]/N,1/2)\pi$.
To see this in greater detail, we plot in Fig.~1b
the zero contours of $E_{n\uparrow}-E_{n\downarrow}$, where $E_{n\sigma}$ are the eigenvalues of the secular matrix.
For this plot, an experimentally relevant value of $H$ is chosen instead.
One can indeed see that the zero
contours for each band all intersect at the points $S$.  On the other hand, these zero contours
are not all confined to being near the zone boundary $k_y \sim \pi/2$.  In particular, zero contours cut across the electron 
pockets, and
as a consequence, even though the two split electron pockets have different areas,
this difference is much smaller than would occur in the longitudinal case.  We have not been
able to identify any symmetry reason for why this occurs.  Plots of the dispersion along the
symmetry axis $k_y=0$ reveal that for the electron band (but not for most of the other bands),
the Zeeman splitting is very small at the bottom of the band,
the so-called $B$ points \cite{Revaz} at $(2n/N,n-1)\pi$.

\begin{figure}
\centerline{\includegraphics[width=3.4in]{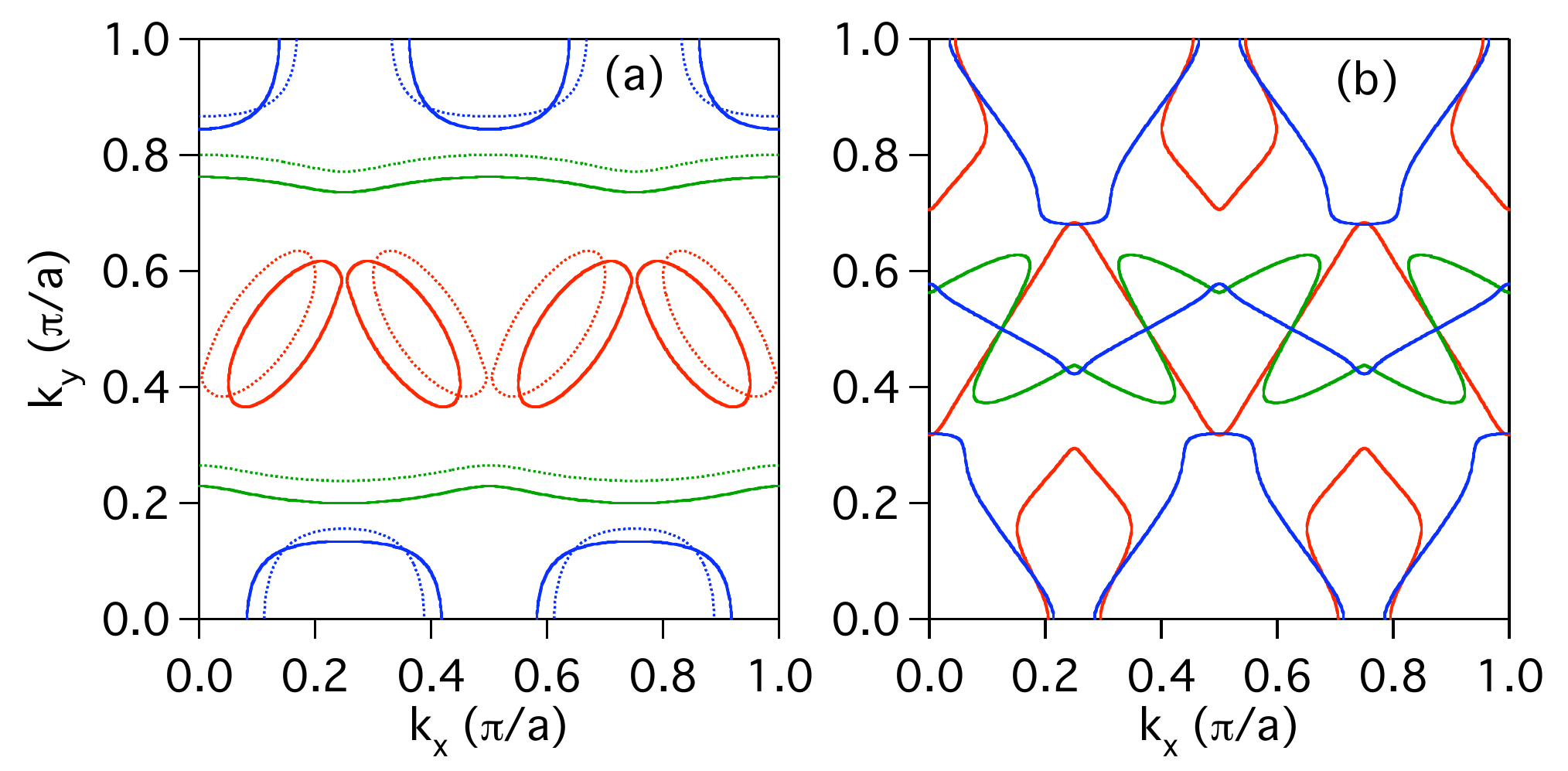}}
\caption{(Color online) $N$=8 transverse case with $V_s$=0.178 eV, $V_c$=0.  (a) Fermi surfaces
for $\frac{g}{2} H$=1000 Tesla (this is artificially large so as to emphasize the Zeeman splitting). 
Solid and dotted curves represent the Fermi surfaces from the `up' and `down' spin bands.
(b) zero contours of $E_{n\uparrow}-E_{n\downarrow}$, where $E_{n\sigma}$ are the eigenvalues of the secular matrix 
with $n$ the band index and $\sigma$ the spin. Here, $\frac{g}{2} H$ is taken
to be 60 Tesla.}
\label{fig1}
\end{figure}

We contrast this with the $N$=10 case shown in Fig.~2.  The fundamental difference here is that
in the paramagnetic case, there are now {\it two} hole bands, each with the same pocket area.
Again, the hole pockets are split by the Zeeman term, maintaining
equal areas.  But interestingly, all bands have a zero contour for  $E_{n\uparrow}-E_{n\downarrow}$ equal to the 
$k_y=\pi/2$ line.  Related to this is the fact that the paramagnetic orbit centers 
of the two
hole pockets (and related cases with $N$ differing by 4) are not
staggered like for the $N$=8 case, but are degenerate (again, this degeneracy is connected to the
presence of two hole bands as compared to the single hole band of the $N$=8 case).
These points are of the form $(4n/N,1/2)\pi$.
Interestingly, the degeneracy present along this horizontal line ($k_y=\pi/2$) does not occur
along any vertical lines.

\begin{figure}
\centerline{\includegraphics[width=3.4in]{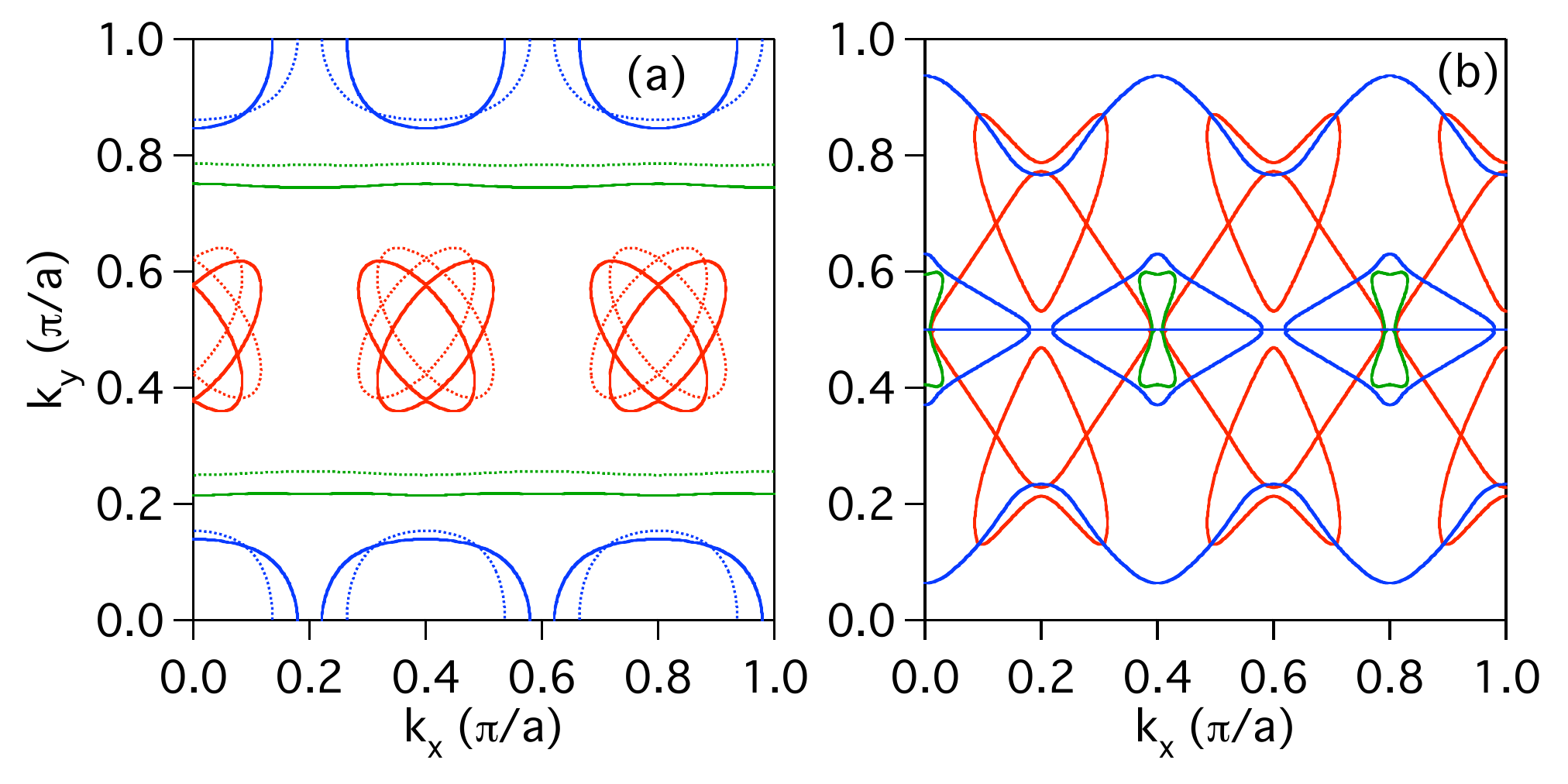}}
\caption{(Color online) $N$=10 transverse case with $V_s$=0.204 eV, $V_c$=0.  (a) Fermi surfaces
for $\frac{g}{2} H$=1000 Tesla (this is artificially large so as to emphasize the Zeeman splitting).
Solid and dotted curves represent the Fermi surfaces from the `up' and `down' spin bands.
(b) zero contours of $E_{n\uparrow}-E_{n\downarrow}$, where $E_{n\sigma}$ are the eigenvalues of the 
secular matrix with $n$ the band index and $\sigma$ the spin.  Here, $\frac{g}{2} H$ is taken
to be 60 Tesla.}
\label{fig2}
\end{figure}

We now calculate the Zeeman splitting of the electron pocket for the $N$=8 and 10 cases, both for
the longitudinal and transverse configurations.  $V_s$ was chosen
to yield an orbit frequency (without Zeeman splitting) for the electron pocket near 530 Tesla as
observed experimentally.\cite{Lifshitz}  As we use a tight binding fit to band theory
for $\epsilon_k$,\cite{Ole} we also ran these calculations with a mass renormalization, $Z$,
of 3 to come into better agreement with the observed cyclotron mass.  This renormalization
is imposed not only on $\epsilon_k$, but on $V_s$ and $V_c$ as well to maintain the pocket size.
Obviously, the Zeeman effect is amplified as $Z$ is increased.  Our results are summarized in Table I.
There, it can be seen that the Zeeman splitting is drastically reduced in the transverse case,
by factors which range from 7.5 to 20
relative to the longitudinal case where full Zeeman splitting is observed. 
We chose to do the calculations for a field of 60 Tesla
to maximize the Zeeman effect for physically relevant fields.  For the $N$=8 case in the transverse
configuration, we have verified that the frequency splitting of the orbits is linear in $H$ up to 60 Tesla.

\begin{table}
\caption{Splitting of the electron orbit frequencies for an applied field, $\frac{g}{2}H$, of 60 Tesla.
$N$ is the stripe period, $Z$ the mass renormalization, and $P$ the polarization of
the spins ($L$ for the spins aligned with the field, $T$ for the spins perpendicular to the field).
The quantity $2\Delta F/H$, where $\Delta F$ is the frequency splitting,
is equivalent to $g m_s/m$ where $g$ is the g-factor
and $m_s$ the `spin' mass.
For $N$=8, $V_s$ is 0.178 eV; for $N$=10, $V_s$ is 0.204 eV.\cite{Lifshitz}
}
\begin{ruledtabular}
\begin{tabular}{cccc}
N & Z & P & $2\Delta F/H$ \\
\colrule
 8 &    1  &  L  & 1.19 \\
    &    &      T   & 0.06 \\
    &    3    &   L  & 3.55 \\
    &     &     T  &  0.19 \\
10 &   1  &  L &  1.30 \\
      &   &      T & 0.17 \\
      &  3   &  L &  3.90 \\
      &   &      T  & 0.52 \\
\end{tabular}
\end{ruledtabular}
\end{table}

To compare to experiment, we need to estimate the effect the splitting has on the oscillation
amplitude.  Following Lifshitz and Kosevich,\cite{Shoenberg} one finds that the Zeeman reduction
in the amplitude, $R_s$, can be written as $\cos(p\pi\Delta F/[H \cos\theta])$ where $p$ is
the harmonic index,
$\Delta F$ the difference of the Zeeman split orbit areas (expressed as a cyclotron
frequency), and $\theta$ the angle of the field relative to the $c$-axis.  Comparing to the
standard formula for $R_s$, we then equate $2\Delta F/H$ with $g m_s/m$ where $m_s$ is
the `spin' mass.  This is the quantity typically reported by experiment.

In Figs.~3 and 4, we show $R_s$ for the electron pocket 
for the $N$=8 and 10 cases, for $Z$=1 and 3, and also for
both longitudinal and transverse configurations (for the fundamental harmonic $p$=1). 
For the transverse case, spin zeros typically occur at very high angles due to the drastically
reduced Zeeman splitting in this case.  This can be contrasted with the longitudinal case,
where full Zeeman splitting occurs resulting in values comparable to a paramagnetic state.
We remind that for the hole pockets, $R_s$ would be identically 1 for the transverse case.

\begin{figure}
\centerline{\includegraphics[width=3.4in]{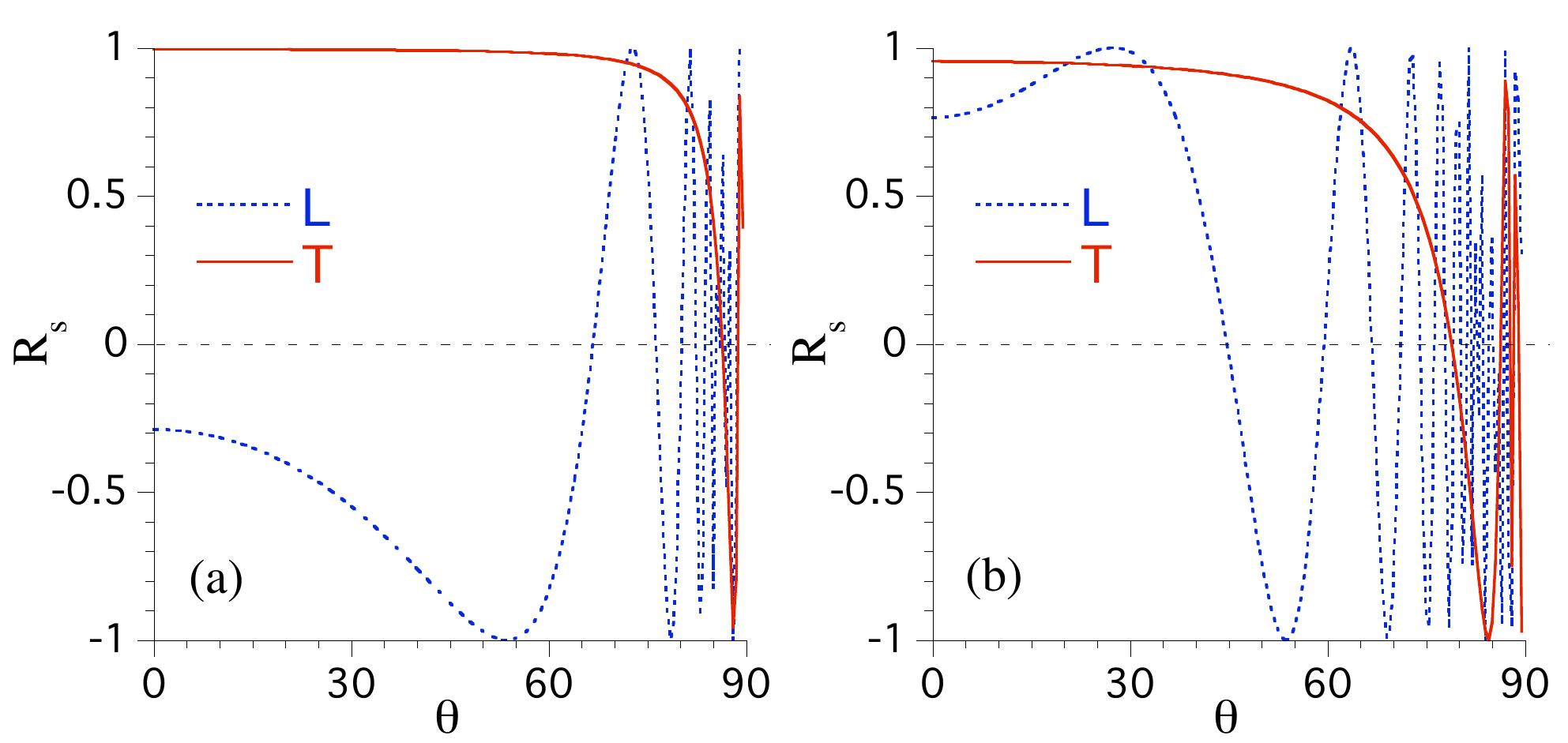}}
\caption{(Color online) Spin amplitude factor $R_s$ versus angle of the field relative to the c-axis
with $\frac{g}{2} H$ = 60 Tesla and $N$=8 for longitudinal ($L$) and transverse ($T$) spins.  (a)
$Z$=1 and (b) $Z$=3, where $Z$ is the mass renormalization.  Same parameters as Fig.~1b.}
\label{fig3}
\end{figure}

\begin{figure}
\centerline{\includegraphics[width=3.4in]{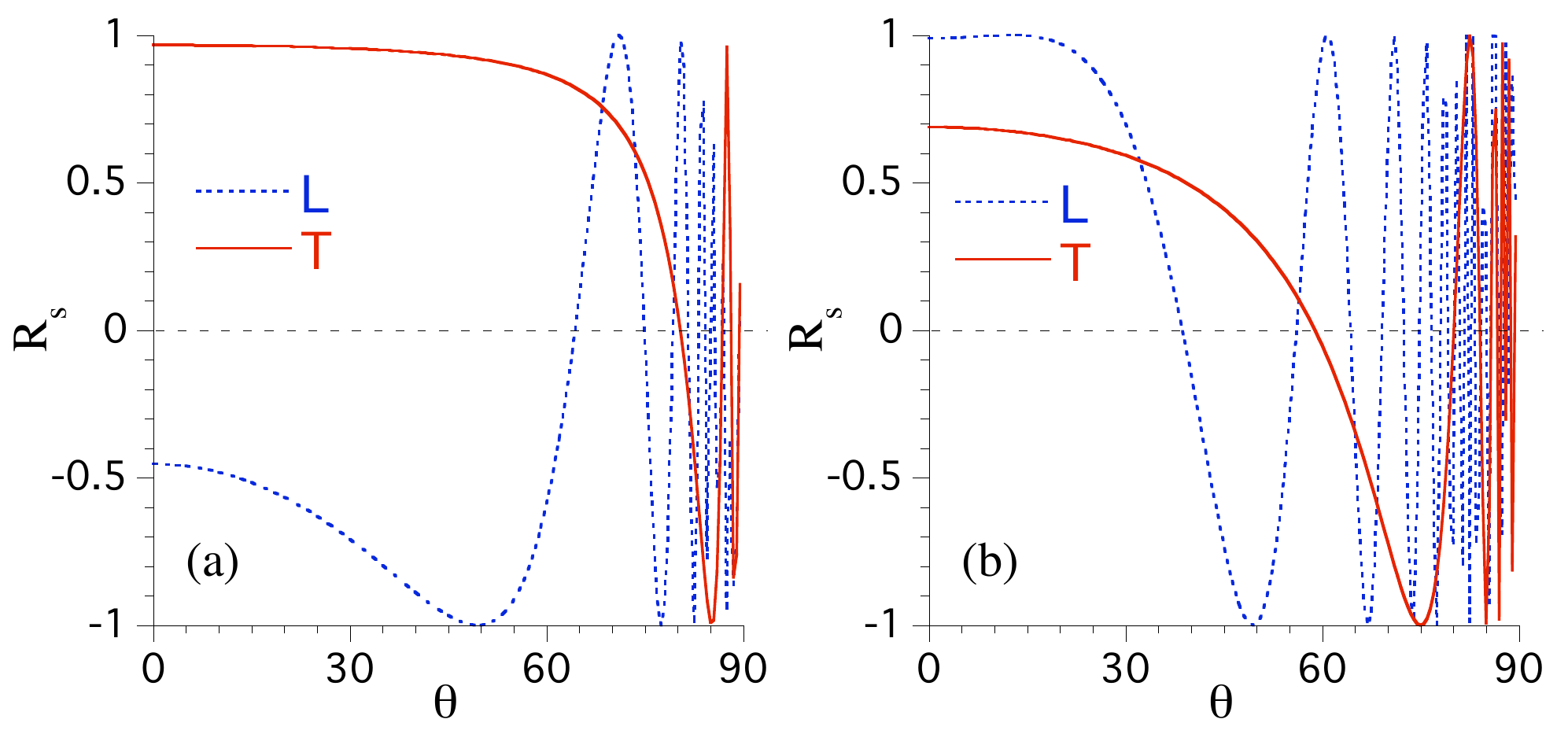}}
\caption{(Color online) Spin amplitude factor $R_s$ versus angle of the field relative to the c-axis
with $\frac{g}{2} H$ = 60 Tesla and $N$=10 for longitudinal ($L$) and transverse ($T$) spins.  (a)
$Z$=1 and (b) $Z$=3, where $Z$ is the mass renormalization.  Same parameters as Fig.~2b.}
\label{fig4}
\end{figure}

We now turn to comparing our results to experiment.  Sebastian \etal \cite{SSPRL} find no spin
zeros up to angles near 60$^\circ$, nor any evidence for a $\pi$ phase shift of the oscillations
that would be consistent with passing through such a zero.
This would definitely be consistent with a hole pocket as
commented on previously by Ramazashvili,\cite{Revaz} but could be consistent with
electron pockets as well.  To know for sure would require extending the measurements to higher
angles, or analyzing higher harmonics.\cite{SS-private}  Unfortunately, this is a quantitative issue,
since we have shown one case in Table I where $2 \Delta F/H$ is as small as 0.06, and this could be
potentially smaller for different choices of $V_s$ and $V_c$.

In contrast, Ramshaw \etal \cite{Ramshaw} were motivated to include an $R_s$ factor in their
fits in an attempt to describe the non-trivial dependence of the overall oscillation amplitude on the
field angle.  From their fits, they
claim zeros at 50$^\circ$ ($g m_s/m$=3.2)
for the fundamental frequency
and 40$^\circ$ ($g m_s/m$=2.1) for a smaller secondary frequency.  For the fundamental frequency, $R_s$ initially
increases as $\theta$ increases from 0.  This would only be consistent with a paramagnetic state, 
or a density wave state of non-magnetic origin, such as the $d$-density wave state.\cite{Sudip} It is also consistent 
with the spin density wave scenario if we assume that the staggered moment has a large component along the field direction.

Noting that these two measurements are different (Ramshaw \etal~measure the c-axis resistance,
whereas Sebastian \etal~measure the change in the skin depth),
the source of the discrepancy between the two has to do with the complex
waveform of the oscillations.  Even if there were only one pocket present, up to four frequencies
could be realized due to (1) bilayer splitting and (2) warping of each individual cylinder (leading
to extremal neck and belly frequencies).  In the analysis of Ref.~\onlinecite{SSPRL}, the authors
did not detect any amplitude suppression or phase shift that would be consistent with the presence
of a spin zero.  They verified this in more extensive work where they also did not detect any such effect in
the second harmonic where it would have been more obvious.\cite{SSPRB}  On the other hand,
Ramshaw \etal \cite{Ramshaw} claim that the presence of closely spaced frequencies can mask
the presence of zeros for each individual frequency.  Their analysis involved a fit including both
the primary and a lower secondary frequency, though they did not include the higher secondary
frequency claimed in other measurements.\cite{Audouard,SSPRB}

To make further progress would require detailed fits including all three frequencies over a wider
field range.  Moreover, these measurements should be supplemented by NMR, neutron, and
magneto-resistance data to look for the orientation of the spins in the high field state.  So far,
little is known about this for underdoped YBa$_2$Cu$_3$O$_{y}$ samples in the doping range magnetic
oscillations have been observed.  For a lower doping, a field induced elastic signal has been
observed by neutrons,\cite{Haug} but the orientation of the spins relative to the field has not
been determined.
Several studies, though, have been made on 1/8 doped La$_{2-x}$Ba$_x$CuO$_4$, where a magnetic stripe state
is stable at zero field.  What is observed from bulk susceptibility studies is a spin flop field of
about 6 Tesla, above which the spins appear to be rotated transverse to the applied field.\cite{Hucker}
Based on this, we feel it quite likely for the extremely high fields used in the magnetic oscillation
studies, the spins are indeed in the transverse configuration.  But further studies will be necessary
to see whether this is indeed the case for YBa$_2$Cu$_3$O$_{y}$ with $y \sim 6.5$.
In the quantum oscillation context,
this could be tested by comparing angle sweeps in the $ac$ and $bc$ planes,
assuming the spins are locked to a particular axis in the plane as occurs for 1/8 doped La$_{2-x}$Ba$_x$CuO$_4$
at low fields.
We note that Sebastian \etal~did not find any significant changes of their
data as a function of the in-plane field angle.\cite{SSPRB} 

We thank Revaz Ramazashvili, Andy Millis, Cyril Proust, Brad Ramshaw, Neil Harrison
and Suchitra Sebastian for 
extensive discussions concerning the spin zeros.  
We are particularly indebted to Suchitra Sebastian for sharing
with us some of her unpublished analysis with Gil Lonzarich.
This work was supported by the US DOE, Office of Science, under contract 
DE-AC02-06CH11357 and by the Center for Emergent Superconductivity, an Energy Frontier Research
Center funded by the US DOE, Office of Science, under Award No.~DE-AC0298CH1088.

\end{document}